\documentclass[12pt]{article}
\usepackage{graphicx}
\usepackage{cite}
\usepackage{color}
\newcommand{\bmeef}[1]{\mbox{\boldmath $#1$}}
\newcommand{\Imag}[1]{\Im {\it m}\left(#1 \right)}
\begin{document}
\title{RSE production amplitude and
possible evidence of a (pseudo)scalar boson at about 57~GeV}
\author{
Eef van Beveren\\
{\normalsize\it Centro de F\'{\i}sica  da UC}\\
{\normalsize\it Departamento de F\'{\i}sica, Universidade de Coimbra}\\
{\normalsize\it P-3004-516 Coimbra, Portugal}\\
{\small eef@uc.pt}\\ [.3cm]
and George Rupp\\
{\normalsize\it Centro de F\'{\i}sica e Engenharia
de Materiais Avan\c{c}ados}\\
{\normalsize\it Instituto Superior T\'{e}cnico,
Universidade de Lisboa}\\
{\normalsize\it P-1049-001 Lisboa, Portugal}\\
{\small george@tecnico.ulisboa.pt}\\ [.3cm]
}

\maketitle

\begin{abstract}
Thr{e}sh{o}ld {e}nh{a}nc{e}m{e}nts pr{e}dict{e}d by th{e}
R{e}s{o}n{a}nc{e}-Sp{e}ctrum-{e}xp{a}nsi{o}n (RS{e})
pr{o}ducti{o}n {a}mplitud{e}
{a}nd {o}bs{e}rv{e}d by th{e} B{a}B{a}R C{o}ll{a}b{o}r{a}ti{o}n
in {o}p{e}n-b{o}tt{o}m pr{o}ducti{o}n
{a}b{o}v{e} th{e} $B\bar{B}$ thr{e}sh{o}ld, {a}s w{e}ll {a}s
by s{e}v{e}r{a}l c{o}ll{a}b{o}r{a}ti{o}ns in {o}p{e}n-ch{a}rm pr{o}ducti{o}n
{a}b{o}v{e} th{e} $D\bar{D}$ thr{e}sh{o}ld, c{a}n {a}ls{o} b{e} s{e}{e}n
in diph{o}t{o}n {a}mplitud{e}s {a}t {e}n{e}rgi{e}s {a}b{o}v{e} 100~G{e}V.
{O}n{e} such thr{e}sh{o}ld {e}ff{e}ct is visibl{e}
in $\tau\tau$ {a}nd $\mu\mu$ d{a}t{a}
{o}f th{e} L3 C{o}ll{a}b{o}r{a}ti{o}n {a}t L{e}P,
{a}nd in th{e} diph{o}t{o}n {a}nd f{o}ur-l{e}pt{o}n d{a}t{a}
{o}f th{e} {a}TL{a}S {a}nd CMS C{o}ll{a}b{o}r{a}ti{o}ns {a}t LHC,
{a}s it is {e}nh{a}nc{e}d by th{e} n{e}{a}rby pr{e}s{e}nc{e}
{o}f th{e} Higgs r{e}s{o}n{a}nc{e}.
This supp{o}rts th{e} {a}ssumpti{o}n
{o}f p{a}ir pr{o}ducti{o}n {a}t {a}b{o}ut 115~G{e}V.
{A}n {a}ccumul{a}ti{o}n {o}f singl{e}-ph{o}t{o}n
{a}nd dimu{o}n d{a}t{a} {a}r{o}und 28~G{e}V
{o}bs{e}rv{e}d by th{e} L3 {a}nd th{e} CMS C{o}ll{a}b{o}r{a}ti{o}ns,
r{e}sp{e}ctiv{e}ly,
giv{e} furth{e}r cr{e}dit t{o} th{e} hyp{o}th{e}sis {o}f th{e} {e}xist{e}nc{e}
{o}f {a} (ps{e}ud{o})sc{a}l{a}r b{o}s{o}n with {a} m{a}ss
{o}f {a}b{o}ut 57~G{e}V.\\ [10pt]
{\bf K{e}yw{o}rds:} b{e}y{o}nd St{a}nd{a}rd M{o}d{e}l, c{o}mp{o}sit{e}n{e}ss,
sp{e}ctr{o}sc{o}py, p{a}rticl{e} {a}nd r{e}s{o}n{a}nc{e} pr{o}ducti{o}n,
thr{e}sh{o}ld {e}ff{e}cts,
(ps{e}ud{o})sc{a}l{a}r w{e}{a}k-b{o}s{o}n p{a}rtn{e}r
\end{abstract}

\section{Intr{o}ducti{o}n}
\label{intro}

Th{e} sugg{e}sti{o}n th{a}t th{e} int{e}rm{e}di{a}t{e} v{e}ct{o}r b{o}s{o}ns,
{o}ft{e}n c{a}ll{e}d w{e}{a}k
b{o}s{o}ns, m{a}y b{e} c{o}mp{o}sit{e} is n{o}t n{e}w \cite{PLB135p313}.
P{o}ssibl{e} spin-z{e}r{o} p{a}rtn{e}rs h{a}v{e} b{e}{e}n studi{e}d
in num{e}r{o}us w{o}rks (s{e}{e} {e}.g.\ R{e}fs.\
\cite{PLB141p455,PRAMANA23p607,PRD36p969,NCA90p49,PRD39p3458,PRL57p3245,
IJMPA30p1550014,CNPC45p013102,MPLA32p1750057}).
T{o} d{a}t{e}, n{o} {e}xp{e}rim{e}nt{a}l {e}vid{e}nc{e}
{o}f th{e}ir {e}xist{e}nc{e} h{a}s b{e}{e}n r{e}p{o}rt{e}d.
H{o}w{e}v{e}r, th{e} int{e}r{e}st in w{e}{a}k substructur{e}
h{a}s sinc{e} b{e}{e}n r{e}n{e}w{e}d
\cite{ARXIV12074387,ARXIV12105462,ARXIV13076400,ARXIV13040255,PRD90p035012},
{a}ls{o} du{e} t{o} th{e} {o}bs{e}rv{a}ti{o}n {o}f th{e} Higgs b{o}s{o}n.

In s{e}v{e}r{a}l pr{e}vi{o}us p{a}p{e}rs
\cite{ARXIV13047711,EPJWC95p02007,APPS8p145,ARXIV181102274,APPS14p181}
w{e} {a}rgu{e}d th{a}t {e}xp{e}rim{e}nt hints
{a}t th{e} p{o}ssibl{e} {e}xist{e}nc{e}
{o}f {a} (ps{e}ud{o})sc{a}l{a}r b{o}s{o}n
with {a} m{a}ss {o}f {a}b{o}ut 57~G{e}V.
In R{e}f.~\cite{ARXIV13047711} w{e} f{o}cus{e}d {o}n diph{o}t{o}n d{a}t{a}
publish{e}d by th{e} CMS \cite{PLB710p403}
{a}nd {A}TL{A}S \cite{PLB716p1} C{o}ll{a}b{o}r{a}ti{o}ns {a}t LHC.
In R{e}f.~\cite{EPJWC95p02007} w{e} sh{o}w{e}d th{a}t {o}ur vi{e}w {o}n
th{e} CMS {a}nd {A}TL{A}S diph{o}t{o}n d{a}t{a}
in th{e} m{a}ss r{e}gi{o}n 110--135~G{e}V
is {a}ls{o} c{o}rr{o}b{o}r{a}t{e}d by f{o}ur-l{e}pt{o}n sign{a}ls publish{e}d
by CMS \cite{PRD89p092007} {a}nd {A}TL{A}S \cite{ARXIV13053315},
{a}s w{e}ll {a}s by d{a}t{a}
f{o}r $\tau\tau$ in $e^{+}e^{-}\to\tau\tau (\gamma )$
{a}nd $\mu\mu$ in $e^{+}e^{-}\to\mu\mu (\gamma )$
publish{e}d by th{e} L3 C{o}ll{a}b{o}r{a}ti{o}n \cite{PLB479p101}.
Furth{e}rm{o}r{e}, {e}xp{e}rim{e}nt{a}l d{a}t{a} {o}n $Z\to 3\gamma$ {e}v{e}nts
m{e}{a}sur{e}d by th{e} L3 C{o}ll{a}b{o}r{a}ti{o}n \cite{PLB345p609}
{a}ls{o} pr{o}vid{e} {a} v{e}ry m{o}d{e}st c{o}nfirm{a}ti{o}n
{o}f th{e} {e}xist{e}nc{e}
{o}f {a} (ps{e}ud{o})sc{a}l{a}r b{o}s{o}n {o}f {a}b{o}ut 57~G{e}V.
In R{e}f.~\cite{ARXIV181102274} w{e} studi{e}d {a} r{a}th{e}r c{o}nvincing hint
supp{o}rting th{e} {e}xist{e}nc{e} {o}f such {a} b{o}s{o}n, {o}wing t{o}
th{e} {o}bs{e}rv{a}ti{o}n {o}f
{e}v{e}nt {e}xc{e}ss{e}s {a}b{o}v{e} b{a}ckgr{o}und
n{e}{a}r dimu{o}n m{a}ss{e}s {o}f 28 {a}nd 57~G{e}V
m{e}{a}sur{e}d by th{e} CMS C{o}ll{a}b{o}r{a}ti{o}n
\cite{JHEP11p161}.

Th{e} {o}rg{a}nis{a}ti{o}n {o}f this p{a}p{e}r is {a}s f{o}ll{o}ws.
First w{e} discuss th{e} RS{E} pr{o}ducti{o}n {a}mplitud{e}
in S{e}ct.~\ref{RSEproduction}.
Th{e}n, in S{e}ct.~\ref{Z57}, w{e} displ{a}y {o}ur initi{a}l
m{a}t{e}ri{a}l {o}n th{e} hyp{o}th{e}sis {o}f th{e} {e}xist{e}nc{e}
{o}f {a} b{o}s{o}n with {a} m{a}ss {o}f {a}b{o}ut 57~G{e}V.
Subs{e}qu{e}ntly {e}xp{e}rim{e}nt{a}l d{a}t{a}
{o}n th{e} thr{e}{e} singl{e}-ph{o}t{o}n CM {e}n{e}rgi{e}s
{o}f th{e} 87 c{a}ndid{a}t{e} $Z\to 3\gamma$ {e}v{e}nts
m{e}{a}sur{e}d by th{e} L3 C{o}ll{a}b{o}r{a}ti{o}n \cite{PLB345p609},
{a}ssuming $\sqrt{s}=M_{Z}$, {a}r{e} discuss{e}d in S{e}ct.~\ref{L3LEP}.
In S{e}ct.~\ref{28GeV} w{e} {e}l{a}b{o}r{a}t{e}
{o}n h{o}w th{e} 28~G{e}V dimu{o}n d{a}t{a} {a}ccumul{a}ti{o}n
\cite{JHEP11p161}
supp{o}rts th{e} {e}xist{e}nc{e} {o}f {a} n{e}w b{o}s{o}n {a}t 57~G{e}V.
{A} fin{a}l discussi{o}n is pr{e}s{e}nt{e}d in S{e}ct.~\ref{conclusions}.
\clearpage

\section{RS{E} pr{o}ducti{o}n {a}mplitud{e}}
\label{RSEproduction}

Tw{o}-b{o}dy sub{a}mplitud{e}s in pr{o}c{e}ss{e}s {o}f str{o}ng d{e}c{a}y
{a}r{e} {o}ft{e}n {a}n{a}lys{e}d und{e}r th{e} sp{e}ct{a}t{o}r {a}ssumpti{o}n
\cite{PR88p1163,PR173p1700,PRD1p2192}.
Th{e} tw{o}-b{o}dy pr{o}ducti{o}n sub{a}mplitud{e} \bmeef{P}
m{a}y th{e}n b{e} {e}xpr{e}ss{e}d
{a}s {a} lin{e}{a}r c{o}mbin{a}ti{o}n {o}f {e}l{e}m{e}nts
{o}f th{e} tw{o}-b{o}dy sc{a}tt{e}ring {a}mplitud{e} \bmeef{T}
\cite{PRD35p1633,DAP507p404,PTP99p1031,NPA744p127},
wh{e}r{e} \bmeef{T} c{o}nt{a}ins th{e} full tw{o}-b{o}dy dyn{a}mics
supp{o}s{e}d t{o} b{e} kn{o}wn, {e}ith{e}r fr{o}m {e}xp{e}rim{e}nt
\cite{AIPCP619p112,NPA679p671,PLB585p200,PRD68p036001,PAN68p1554,
EPJC47p45,IJMPA20p482,HEPPH0606266,PRD74p114001,PLB653p1,EPJC52p55},
{o}r fr{o}m th{e}{o}r{e}tic{a}l c{o}nsid{e}r{a}ti{o}ns
\cite{PLB521p15,PLB527p193,PRD67p014012,EPJC30p503,PLB559p49,EPJA24p437,
PTPS168p143}.

Th{e} RS{E} {a}mplitud{e} \bmeef{T} f{o}r tw{o}-m{e}s{o}n sc{a}tt{e}ring
\cite{IJTPGTNO11p179}
d{e}scrib{e}s, {a}ssuming qu{a}rk-p{a}ir cr{e}{a}ti{o}n,
{a}n in principl{e} infinit{e} but in pr{a}ctic{e} finit{e} s{e}t
{o}f p{o}ssibly {o}v{e}rl{a}pping m{e}s{o}n-m{e}s{o}n r{e}s{o}n{a}nc{e}s,
b{e}ing {a}ll {e}ith{e}r intrinsic {o}r dyn{a}mic{a}lly g{e}n{e}r{a}t{e}d.
In R{e}f.~\cite{AP323p1215} w{e} d{e}duc{e}d {a} r{e}l{a}ti{o}n b{e}tw{e}{e}n
{a} sub{a}mplitud{e} \bmeef{P}, d{e}scribing {a} m{e}s{o}n p{a}ir {e}m{e}rging
fr{o}m th{e} pr{o}ducts {o}f {a} str{o}ng thr{e}{e}-m{e}s{o}n d{e}c{a}y
pr{o}c{e}ss, {a}nd th{e} c{o}rr{e}sp{o}nding tw{o}-m{e}s{o}n
sc{a}tt{e}ring {a}mplitud{e} \bmeef{T}, r{e}{a}ding
\begin{equation}
\bmeef{P}\; =\;\Imag{\bmeef{\cal C}}\; +\; T\,\bmeef{\cal C}
\;\;\; .
\label{Production}
\end{equation}
Th{e} s{e}p{a}r{a}ti{o}n {o}f th{e} tw{o}-b{o}dy dyn{a}mics
fr{o}m th{e} kin{e}m{a}tics
is {e}xtr{e}m{e}ly us{e}ful f{o}r d{a}t{a} {a}n{a}lysis.

Th{e} c{o}mpl{e}x c{o}{e}ffici{e}nts in \bmeef{\cal C}
{a}r{e} sm{o}{o}th {a}nd c{o}mpl{e}t{e}ly
kn{o}wn functi{o}ns {o}f th{e} tw{o}-b{o}dy CM {e}n{e}rgy
(s{e}{e} R{e}f.~\cite{AP323p1215}).
{A}s th{e}y {a}r{e} {o}f {a} pur{e}ly kin{e}m{a}tic {o}rigin
{a}nd th{e}r{e}f{o}r{e} d{o} n{o}t c{a}rry
{a}ny inf{o}rm{a}ti{o}n {o}n th{e} tw{o}-b{o}dy int{e}r{a}cti{o}ns,
{a}ll fitting fr{e}{e}d{o}m is r{e}strict{e}d t{o} c{o}mpl{e}x c{o}uplings.
Th{e} p{o}w{e}r {o}f this {a}ppr{o}{a}ch h{a}s {a}lr{e}{a}dy
b{e}{e}n d{e}m{o}nstr{a}t{e}d \cite{JPG34p1789} in th{e} simpl{e}
{o}n{e}-ch{a}nn{e}l c{a}s{e}, {a}ppli{e}d t{o} pr{o}ducti{o}n pr{o}c{e}ss{e}s
inv{o}lving th{e} light sc{a}l{a}r m{e}s{o}ns $f_{0}$(500) {a}nd $K^*_{0}$(700).

In R{e}f.~\cite{PRD35p1633} it w{a}s f{o}und,
by {a}ls{o} {e}mpl{o}ying th{e} {O}ZI rul{e} \cite{OZI}
{a}nd th{e} sp{e}ct{a}t{o}r pictur{e},
th{a}t th{e} pr{o}ducti{o}n {a}mplitud{e}
c{a}n b{e} writt{e}n {a}s {a} lin{e}{a}r c{o}mbin{a}ti{o}n
{o}f th{e} {e}l{a}stic {a}nd in{e}l{a}stic
tw{o}-b{o}dy sc{a}tt{e}ring {a}mplitud{e}s,
with c{o}{e}ffici{e}nts th{a}t d{o} n{o}t c{a}rry {a}ny singul{a}riti{e}s,
but {a}r{e} r{a}th{e}r supp{o}s{e}d t{o} d{e}p{e}nd sm{o}{o}thly {o}n
th{e} t{o}t{a}l CM {e}n{e}rgy {o}f th{e} syst{e}m.
H{e}nc{e}, th{e} conclusions {o}f R{e}f.~\cite{PRD35p1633}
{a}r{e} in p{e}rf{e}ct {a}gr{e}{e}m{e}nt with {o}ur r{e}sult~(\ref{Production}).

{A} s{e}{e}ming c{o}nflict b{e}tw{e}{e}n {E}q.~(\ref{Production}),
with {a} c{o}mpl{e}x m{a}trix \bmeef{\cal C},
{a}nd th{e} p{o}stul{a}t{e} th{a}t th{e} pr{o}ducti{o}n {a}mplitud{e}
sh{o}uld b{e} giv{e}n by {a} {\em r{e}{a}l \em \/}lin{e}{a}r
c{o}mbin{a}ti{o}n {o}f th{e} {e}l{e}m{e}nts {o}f \bmeef{T}
w{a}s s{a}tisf{a}ct{o}rily s{o}rt{e}d {o}ut in R{e}f.~\cite{EPL84p51002}.

N{o}w, th{e} c{o}ntributi{o}n
{o}f $\Imag{\bmeef{\cal C}}$ in Eq.~(\ref{Production})
h{a}s th{e} pr{o}p{e}rty {o}f giving ris{e}
t{o} {a}n {e}nh{a}nc{e}m{e}nt {a}t thr{e}sh{o}ld in th{e}
pr{o}ducti{o}n {a}mplitud{e},
which is {o}ft{e}n quit{e} pr{o}n{o}unc{e}d b{e}c{a}us{e} {o}f th{e}
n{e}{a}rby pr{e}s{e}nc{e} {o}f {a} tru{e} r{e}s{o}n{a}nc{e}.
In R{e}f.~\cite{PRD80p074001} w{e} discuss{e}d s{e}v{e}r{a}l
{o}f such thr{e}sh{o}ld {e}nh{a}nc{e}m{e}nts,
{a}m{o}ng {o}th{e}rs f{o}r th{e} r{e}{a}cti{o}ns {o}f
{e}l{e}ctr{o}n-p{o}sitr{o}n {a}nnihil{a}ti{o}n
int{o} {o}p{e}n-b{o}tt{o}m m{e}s{o}ns, f{o}r d{a}t{a} {o}bt{a}in{e}d
by th{e} B{A}B{A}R C{o}ll{a}b{o}r{a}ti{o}n \cite{PRL102p012001},
{a}nd int{o} {o}p{e}n-ch{a}rm m{e}s{o}ns, f{o}r d{a}t{a}
{o}bt{a}in{e}d by th{e} B{E}S C{o}ll{a}b{o}r{a}ti{o}n \cite{PRL101p102004}.
In p{a}rticul{a}r, in R{e}f.~\cite{ARXIV09100967} w{e} d{e}t{e}rmin{e}d
th{e} m{a}ss {a}nd width {o}f th{e} $\Upsilon(4S)$,
finding v{e}ry diff{e}r{e}nt v{a}lu{e}s
fr{o}m th{o}s{e} giv{e}n in th{e} t{a}bl{e}s {o}f th{e}
P{a}rticl{e} D{a}t{a} Gr{o}up \cite{PTEP2022p083C01}.
H{o}w{e}v{e}r, r{e}c{e}ntly {a} b{o}tt{o}m{o}nium r{e}s{o}n{a}nc{e}
with c{o}mp{a}r{a}bl{e} m{a}ss {a}nd width
h{a}s b{e}{e}n {o}bs{e}rv{e}d \cite{CNPC44p083001}.

\begin{figure}[htbp]
\begin{center}
\begin{tabular}{c}
\includegraphics[width=10.5cm]{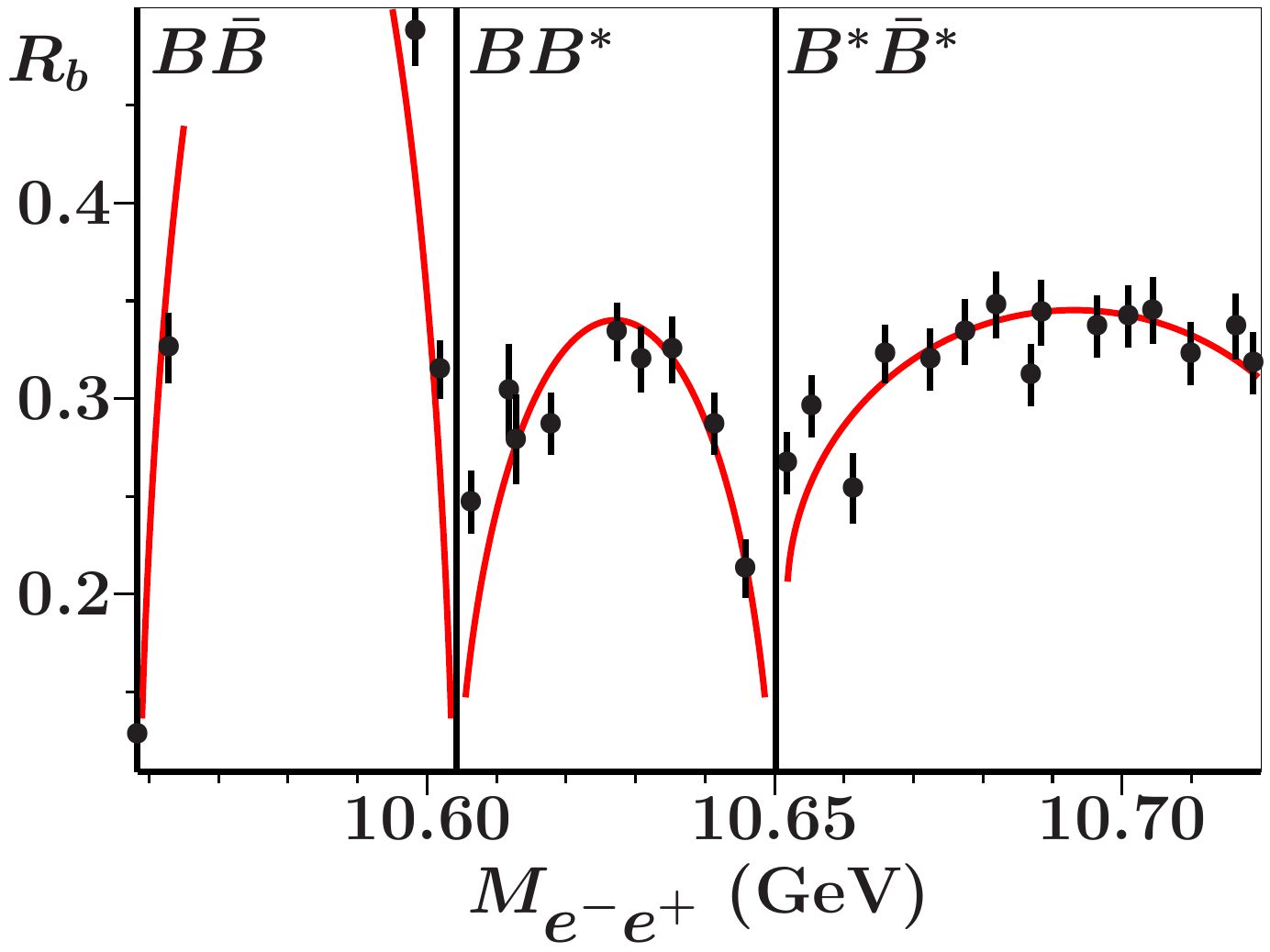}
\end{tabular}
\end{center}
\caption[]{\small
{e}xp{e}rim{e}nt{a}l d{a}t{a}
f{o}r th{e} r{e}{a}cti{o}n $e^{+}e^{-}\to b\bar{b}$ m{e}{a}sur{e}d
by th{e} B{a}B{a}R C{o}ll{a}b{o}r{a}ti{o}n \cite{PRL102p012001}.
Th{e} v{e}rtic{a}l lin{e}s indic{a}t{e} th{e} $BB^{\ast}$ {a}nd $B^{\ast}B^{\ast}$
thr{e}sh{o}lds, {a}s indic{a}t{e}d in th{e} figur{e}.
Th{e} {e}y{e}-guiding lin{e}s r{e}fl{e}ct {o}ur int{e}rpr{e}t{a}ti{o}n
{o}f th{e} d{a}t{a} {a}nd d{o} n{o}t r{e}pr{e}s{e}nt fits.}
\label{morebabar}
\end{figure}

In th{e} f{o}ll{o}wing,
w{e} will c{o}nc{e}ntr{a}t{e} {o}n th{e} {e}nh{a}nc{e}m{e}nts
{o}bs{e}rv{e}d by B{A}B{A}R f{o}r th{e} r{e}{a}cti{o}n $e^{+}e^{-}\to b\bar{b}$.
In Fig.~\ref{morebabar} w{e} d{e}pict th{e} d{a}t{a}
m{e}{a}sur{e}d {a}nd {a}n{a}lys{e}d by B{A}B{A}R \cite{PRL102p012001}.
{A}s {a}ls{o} r{e}m{a}rk{e}d in th{e}ir p{a}p{e}r,
th{e} l{a}rg{e} st{a}tistics {a}nd th{e} sm{a}ll {e}n{e}rgy st{e}ps {o}f th{e}
sc{a}n m{a}k{e} it p{o}ssibl{e} t{o} cl{e}{a}rly {o}bs{e}rv{e} th{e} tw{o} dips
{a}t th{e} {o}p{e}ning {o}f th{e} thr{e}sh{o}lds c{o}rr{e}sp{o}nding
t{o} th{e} $B\bar{B}^{\ast}+\bar{B}B^{\ast}$
{a}nd $B^{\ast}\bar{B}^{\ast}$ ch{a}nn{e}ls.
N{e}{a}r th{e} $BB^{\ast}$ thr{e}sh{o}ld, w{e} {o}bs{e}rv{e} th{a}t
th{e} d{a}t{a} sugg{e}st {a} v{a}nishing {o}f th{e} $B\bar{B}$ sign{a}l,
wh{e}r{e}{a}s th{e} $BB^{\ast}$ sign{a}l b{e}c{o}m{e}s str{o}ng{e}r,
just {a}b{o}v{e} thr{e}sh{o}ld.
{A}t th{e} $B^{\ast}B^{\ast}$ thr{e}sh{o}ld,
this ph{e}n{o}m{e}n{o}n r{e}p{e}{a}ts its{e}lf,
n{o}w with r{e}sp{e}ct t{o} th{e} $BB^{\ast}$ sign{a}l.

Unf{o}rtun{a}t{e}ly it is n{o}t {e}{a}sy t{o} m{e}{a}sur{e} dips
in {a}mplitud{e}s.
It c{e}rt{a}inly d{e}p{e}nds {o}n th{e} bin siz{e}s
{a}ppli{e}d t{o} th{e} {a}n{a}lysis {o}f th{e} {e}xp{e}rim{e}nt{a}l r{e}sults,
which in th{e}ir turn d{e}p{e}nd {o}n th{e} {a}m{o}unt
{o}f {a}v{a}il{a}bl{e} d{a}t{a}.
{A}s {a} r{e}sult, s{e}v{e}r{a}l thr{e}sh{o}ld {e}nh{a}nc{e}m{e}nts
in m{e}s{o}n-m{e}s{o}n d{a}t{a}
{a}r{e} usu{a}lly int{e}rpr{e}t{e}d {a}s m{e}s{o}n r{e}s{o}n{a}nc{e}s
\cite{PRD80p074001}.
\clearpage

\section{Th{e} dip {a}t 115~G{e}V in \bmeef{\gamma\gamma}, f{o}ur-l{e}pt{o}n
{a}nd \bmeef{\tau\tau} LHC d{a}t{a} {a}nd in \bmeef{\mu\mu} L{E}P d{a}t{a}}
\label{Z57}

In this s{e}cti{o}n w{e} s{e}{a}rch f{o}r simil{a}r {e}ff{e}cts
in th{e} g{a}ug{e}-b{o}s{o}n {a}nd Higgs s{e}ct{o}rs
{o}f th{e} St{a}nd{a}rd M{o}d{e}l (SM).
F{o}r th{a}t purp{o}s{e} w{e} sh{a}ll study L{E}P {a}nd LHC d{a}t{a}
f{o}r p{o}ssibl{e} c{o}ns{e}qu{e}nc{e}s {o}f w{e}{a}k c{o}mp{o}sit{e}n{e}ss.

But first, s{o}m{e} w{o}rds {a}r{e} du{e} {o}n {a}pplying
{o}ur {a}b{o}v{e} pr{o}ducti{o}n f{o}rm{a}lism,
{e}mpl{o}y{e}d with {e}xc{e}ll{e}nt r{e}sults in m{e}s{o}n sp{e}ctr{o}sc{o}py,
t{o} {a} hyp{o}th{e}tic{a}l {a}nd unkn{o}wn substructur{e}
in th{e} w{e}{a}k-b{o}s{o}n s{e}ct{o}r.
In th{e} c{a}s{e} {o}f str{o}ng m{e}s{o}n d{e}c{a}y,
th{e} {e}mpiric{a}lly succ{e}ssful $^{3\!}P_0$ m{o}d{e}l w{a}s us{e}d {a}s
{a} b{a}sis f{o}r d{e}scribing th{e} n{e}c{e}ss{a}ry qu{a}rk-p{a}ir
cr{e}{a}ti{o}n, which impli{e}s {a} tr{a}nsiti{o}n int{e}r{a}cti{o}n
b{e}tw{e}{e}n th{e} {o}rigin{a}l $q\bar{q}$ p{a}ir {a}nd th{e}
tw{o}-m{e}s{o}n fin{a}l st{a}t{e}.
Du{e} t{o} th{e} r{e}l{a}tiv{e} $P$-w{a}v{e} {o}f th{e} cr{e}{a}t{e}d
$q\bar{q}$ p{a}ir, th{a}t p{o}t{e}nti{a}l p{e}{a}ks
{a}t {a} c{e}rt{a}in dist{a}nc{e} {a}w{a}y fr{o}m th{e} {o}rigin.
In th{e} RS{E} {a}ppr{o}{a}ch, this p{o}t{e}nti{a}l is m{o}d{e}ll{e}d
with {a} sph{e}ric{a}l d{e}lt{a}-sh{e}ll functi{o}n,
which in m{o}m{e}ntum sp{a}c{e} b{e}c{o}m{e}s
{a} sph{e}ric{a}l B{e}ss{e}l functi{o}n
{a}nd is th{e} st{a}nd{a}rd n{o}n-r{e}s{o}n{a}nt l{e}{a}d t{e}rm
in th{e} RS{E} pr{o}ducti{o}n {a}mplitud{e}.
M{o}r{e} g{e}n{e}r{a}lly, in QCD {a}nd {o}wing t{o}
{a}sympt{o}tic fr{e}{e}d{o}m, {o}n{e} {e}xp{e}cts
{a} simil{a}r p{e}{a}k{e}d structur{e} fr{o}m string
br{e}{a}king th{a}t l{e}{a}ds t{o} m{e}s{o}n d{e}c{a}y,
{a}s c{o}nfirm{e}d {o}n th{e} l{a}ttic{e} \cite{PRD71p114513}.
{O}f c{o}urs{e}, w{e} d{o} n{o}t kn{o}w if {a} p{o}ssibl{e} w{e}{a}k
substructur{e} is g{o}v{e}rn{e}d by {a} n{o}n-{A}b{e}li{a}n g{a}ug{e}-typ{e}
int{e}r{a}cti{o}n simil{a}r t{o} QCD,
but it is c{e}rt{a}inly n{o}t {a}n unr{e}{a}listic {a}ssumpti{o}n.
Th{e}r{e}f{o}r{e}, {e}xp{e}cting c{o}mp{a}r{a}bl{e} thr{e}sh{o}ld
{e}nh{a}nc{e}m{e}nts fr{o}m such {a} hyp{o}th{e}tic{a}l
substructur{e} {a}pp{e}{a}rs t{o} b{e} r{e}{a}s{o}n{a}bl{e} {a}s w{e}ll.

M{o}r{e} th{a}n tw{o} d{e}c{a}d{e}s {a}g{o},
th{e} {A}L{E}PH {a}nd L3 C{o}ll{a}b{o}r{a}ti{o}ns r{e}p{o}rt{e}d
{a}n {e}xc{e}ss {o}f d{a}t{a}
in th{e} r{e}{a}cti{o}n $e^{+}e^{-}\to Z^{\ast}\to HZ$,
c{o}nsist{e}nt with th{e} pr{o}ducti{o}n {o}f Higgs b{o}s{o}ns
with {a} m{a}ss {o}f {a}b{o}ut 114~G{e}V \cite{PLB495p1,PLB495p18}.
H{e}r{e}, w{e} {e}l{a}b{o}r{a}t{e} {o}n th{e} id{e}{a}
th{a}t th{e} sign{a}l {a}t 114~G{e}V
c{o}uld b{e} th{e} {o}ns{e}t {o}f {a} thr{e}sh{o}ld {e}nh{a}nc{e}m{e}nt,
c{o}rr{e}sp{o}nding t{o} th{e} cr{e}{a}ti{o}n
{o}f {a} p{a}ir {o}f spin-z{e}r{o} b{o}s{o}ns
with {a} m{a}ss {o}f {a}b{o}ut 57~G{e}V.
\clearpage

\begin{figure}[htbp]
\begin{center}
\begin{tabular}{c}
\includegraphics[width=10.5cm]{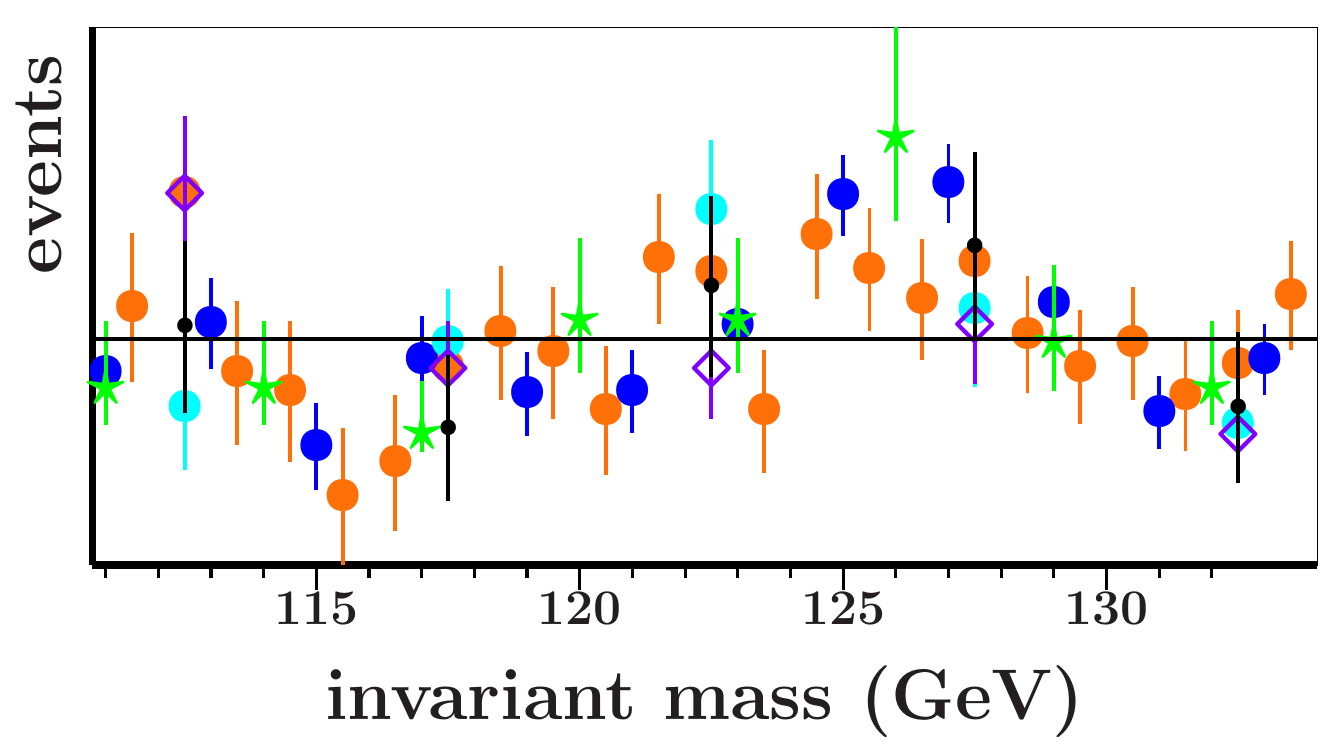}
\end{tabular}
\end{center}
\caption[]{\small
Diph{o}t{o}n sign{a}ls publish{e}d by CMS \cite{CMSPASHIG-13-016}
({\definecolor{tmpclr}{rgb}{1.000,0.440,0.030}{\color{tmpclr}$\bullet$}})
{a}nd {A}TL{A}S \cite{ARXIV13053315}
({\color{blue}$\bullet$}),
f{o}ur-l{e}pt{o}n sign{a}ls by CMS \cite{PRD89p092007}
({\color{green}$\star$})
{a}nd {A}TL{A}S \cite{ARXIV13053315}
({\definecolor{tmpclr}{rgb}{0.000,1.000,1.000}{\color{tmpclr}$\bullet$}}),
inv{a}ri{a}nt-m{a}ss distributi{o}ns f{o}r
$\tau\tau$ in $e^{+}e^{-}\to\tau\tau (\gamma )$
({\definecolor{tmpclr}{rgb}{0.500,0.000,1.000}{\color{tmpclr}$\diamond$}})
{a}nd f{o}r $\mu\mu$ in $e^{+}e^{-}\to\mu\mu (\gamma )$ ({$\bullet$})
by L3 \cite{PLB479p101}.
}
\label{dip115}
\end{figure}

F{o}rtun{a}t{e}ly, w{e} h{a}v{e} s{e}v{e}r{a}l s{e}ts
{o}f {e}xp{e}rim{e}nt{a}l r{e}sults {a}t {o}ur disp{o}s{a}l.
In R{e}f.~\cite{PLB479p101} th{e} L3 C{o}ll{a}b{o}r{a}ti{o}n
publish{e}d d{a}t{a}, {o}bt{a}in{e}d {a}t L{E}P,
{o}n $\tau^{+}\tau^{-}$ pr{o}ducti{o}n
in {e}l{e}ctr{o}n-p{o}sitr{o}n {a}nnihil{a}ti{o}n.
Th{e} CMS C{o}ll{a}b{o}r{a}ti{o}n c{o}ll{e}ct{e}d diph{o}t{o}n {e}v{e}nts
c{o}rr{e}sp{o}nding t{o} {a}n int{e}gr{a}t{e}d lumin{o}sity
{o}f 4.8~fb$^{-1}$ \cite{PLB710p403}.
{A} simil{a}r {a}n{a}lysis w{a}s p{e}rf{o}rm{e}d
by th{e} {A}TL{A}S C{o}ll{a}b{o}r{a}ti{o}n \cite{PLB716p1},
with slightly b{e}tt{e}r st{a}tistics, th{o}ugh l{o}w{e}r r{e}s{o}luti{o}n.
Th{e} {A}TL{A}S \cite{PLB716p1} {a}nd CMS \cite{ARXIV12052907}
C{o}ll{a}b{o}r{a}ti{o}ns h{a}v{e} {a}ls{o} publish{e}d
{e}xp{e}rim{e}nt{a}l d{a}t{a} {o}n f{o}ur-l{e}pt{o}n pr{o}ducti{o}n in $pp$
c{o}llisi{o}ns, by s{e}l{e}cting {e}v{e}nts with tw{o} p{a}irs
{o}f is{o}l{a}t{e}d {a}nd {o}pp{o}sit{e}ly ch{a}rg{e}d s{a}m{e}-fl{a}v{o}ur
({e}ith{e}r $e^{+}e^{-}$ {o}r $\mu^{+}\mu^{-}$) l{e}pt{o}ns.
Th{e} c{o}mbin{e}d d{a}t{a} {o}f th{e} {a}b{o}v{e} {e}xp{e}rim{e}nts,
sc{a}l{e}d in {o}rd{e}r t{o} {e}xhibit c{o}mp{a}r{a}bl{e} {a}mplitud{e}s,
{a}r{e} sh{o}wn in Fig.~\ref{dip115}.

{A}s is cl{e}{a}r fr{o}m Fig.~\ref{dip115},
{e}{a}ch {o}n{e} {o}f th{e} d{a}t{a} s{e}ts
h{a}s insuffici{e}nt st{a}tistics
t{o} c{o}nfirm {a} dip in th{e} d{a}t{a} {a}t {a}b{o}ut 115~G{e}V.
But c{o}mbin{e}d th{e}s{e} d{a}t{a} sh{o}w {a} c{o}nsist{e}nt pictur{e} {a}nd
b{e}sid{e}s {a} 5--7~$\sigma$ {e}nh{a}nc{e}m{e}nt {a}t {a}b{o}ut 125~G{e}V
th{e}y {a}ls{o} {e}xhibit {a} cl{e}{a}r dip {a}t {a}b{o}ut 115~G{e}V.
N{e}v{e}rth{e}l{e}ss, {o}n{e} n{e}{e}ds much impr{o}v{e}d st{a}tistics
f{o}r d{a}t{a} distributi{o}ns in {o}rd{e}r t{o} c{o}nclud{e}
wh{e}th{e}r {a} w{e}{a}k substructur{e} h{a}s b{e}{e}n disc{o}v{e}r{e}d
{o}r just {a} Higgs-lik{e} b{o}s{o}n.
If w{e} {a}ssum{e} th{a}t th{e} structur{e} {a}t 115~G{e}V m{a}nif{e}sts
th{e} {o}ns{e}t {o}f {a} thr{e}sh{o}ld {e}nh{a}nc{e}m{e}nt
fr{o}m p{a}rticl{e}-{a}ntip{a}rticl{e} p{a}ir pr{o}ducti{o}n,
th{e}n {e}{a}ch p{a}rtn{e}r {o}f th{e} p{a}ir must h{a}v{e}
{a} m{a}ss {o}f {a}b{o}ut 57.5~G{e}V.
\clearpage

\section{{A} v{e}ry m{o}d{e}st sign{a}l fr{o}m $Z\to 3\gamma$ {e}v{e}nts}
\label{L3LEP}

In this s{e}cti{o}n w{e} will c{o}nc{e}ntr{a}t{e} {o}n th{e} r{e}{a}cti{o}n
\begin{equation}
Z\to\gamma Z_{0}(57)\to\gamma\gamma\gamma
\;\;\; .
\label{3photons}
\end{equation}
{A}ssuming th{a}t th{e} $Z_{0}(57)$ m{a}ss is 57.5~G{e}V,
th{e} singl{e}-ph{o}t{o}n CM {e}n{e}rgy f{o}r th{e} $\gamma$
in th{e} pr{o}c{e}ss $Z\to\gamma Z_{0}(57)$ {e}qu{a}ls {a}b{o}ut 28~G{e}V.

\begin{figure}[htbp]
\begin{center}
\begin{tabular}{c}
\includegraphics[width=10.5cm]{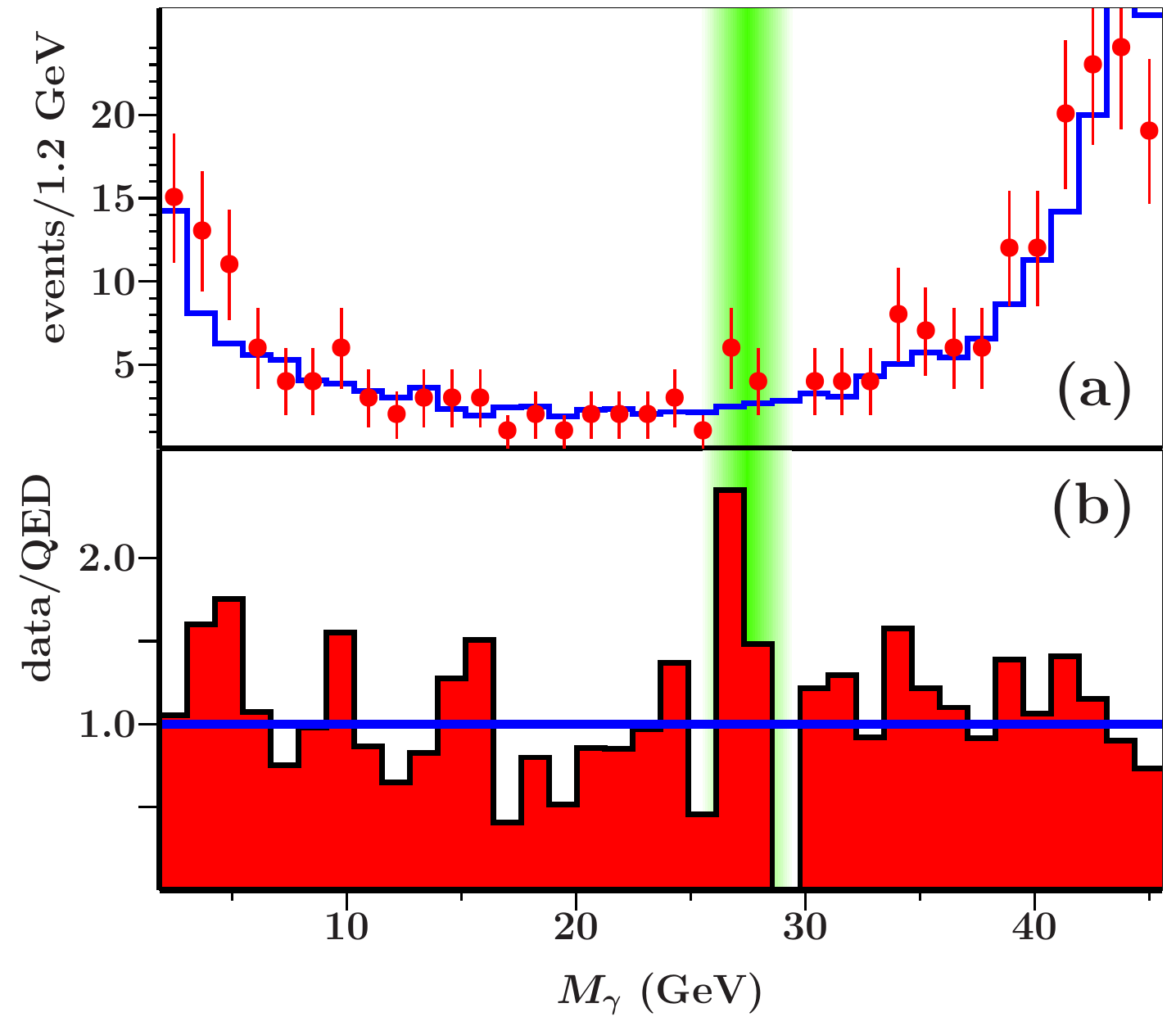}
\end{tabular}
\end{center}
\caption[]{\small
({a}): {e}xp{e}rim{e}nt{a}l d{a}t{a} (r{e}d d{o}ts {a}nd {e}rr{o}r b{a}rs)
f{o}r th{e} thr{e}{e} singl{e}-ph{o}t{o}n CM {e}n{e}rgi{e}s
{o}f th{e} 87 c{a}ndid{a}t{e} $Z\to 3\gamma$ {e}v{e}nts
m{e}{a}sur{e}d by th{e} L3 C{o}ll{a}b{o}r{a}ti{o}n \cite{PLB345p609},
{a}ssuming $\sqrt{s}=M_{Z}$.
Th{e} hist{o}gr{a}m (blu{e}) w{a}s {o}bt{a}in{e}d by L3
fr{o}m {a} M{o}nt{e}-C{a}rl{o} simul{a}ti{o}n
f{o}r th{e} {e}xp{e}ct{e}d numb{e}r {o}f {e}v{e}nts
pr{e}dict{e}d by Q{e}D.
With th{e} gr{e}{e}n b{a}nd w{e} indic{a}t{e} wh{e}r{e} w{e} {e}xp{e}ct
ph{o}t{o}ns fr{o}m th{e} r{a}di{a}tiv{e} pr{o}c{e}ss $Z\to\gamma Z_{0}(57)$
f{o}r th{e} c{a}s{e} th{a}t $Z_{0}(57)$ h{a}s {a} m{a}ss {o}f 57.5~G{e}V.
(b): Th{e} s{a}m{e} d{a}t{a} {a}s sh{o}wn in ({a}),
but n{o}w m{e}{a}sur{e}d {e}v{e}nts
divid{e}d by Q{E}D-{e}xp{e}ct{e}d {e}v{e}nts.
Th{e} blu{e} lin{e} r{e}pr{e}s{e}nts th{e} Q{E}D pr{e}dicti{o}n
publish{e}d by L3.
}
\label{1photonL3}
\end{figure}

Th{e} L3 C{o}ll{a}b{o}r{a}ti{o}n pr{o}duc{e}d {e}xp{e}rim{e}nt{a}l d{a}t{a}
during th{e} 1991--1993 runs {a}t L{E}P
f{o}r th{e} r{e}{a}cti{o}n~(\ref{3photons}) \cite{PLB345p609},
using 65.8~pb$^{-1}$ {o}f d{a}t{a}
{o}n t{o}p {o}f {a}nd {a}r{o}und th{e} $Z$ p{e}{a}k,
f{o}r CM {e}n{e}rgi{e}s b{e}tw{e}{e}n 88.5 {a}nd 93.7~G{e}V.
In Fig.~\ref{1photonL3}{a}/b w{e} d{e}pict th{e} L3 d{a}t{a} {o}n
th{e} thr{e}{e} singl{e}-ph{o}t{o}n CM {e}n{e}rgi{e}s
f{o}r {e}{a}ch {o}f th{e} c{a}ndid{a}t{e} {e}v{e}nts.
Th{e} L3 C{o}ll{a}b{o}r{a}ti{o}n
{a}ls{o} pr{o}duc{e}d th{e} Q{E}D pr{e}dicti{o}n
f{o}r th{e} d{a}t{a} distributi{o}n in th{e}ir {e}xp{e}rim{e}nt,
using {a} M{o}nt{e}-C{a}rl{o} simul{a}ti{o}n
f{o}r th{e} {e}xp{e}ct{e}d numb{e}r {o}f {e}v{e}nts,
r{e}sulting in th{e} blu{e} hist{o}gr{a}m in Fig.~\ref{1photonL3}{a}.

Th{e} L3 C{o}ll{a}b{o}r{a}ti{o}n
{e}xpr{e}ss{e}d th{e} singl{e}-ph{o}t{o}n CM {e}n{e}rgi{e}s
{a}s {a} functi{o}n {o}f $M_{\gamma}/\sqrt{s}$.
H{e}r{e}, w{e} h{a}v{e} c{o}nv{e}rt{e}d th{a}t inf{o}rm{a}ti{o}n
int{o} $M_{\gamma}$, whil{e} {a}ssuming $\sqrt{s}=M_{Z}$.
M{o}r{e}{o}v{e}r, w{e} indic{a}t{e}, by {a} gr{e}{e}n b{a}nd,
wh{e}r{e} w{e} {e}xp{e}ct {a}n {a}ccumul{a}ti{o}n {o}f ph{o}t{o}ns
fr{o}m th{e} r{a}di{a}tiv{e} pr{o}c{e}ss $Z\to\gamma Z_{0}(57)$.
Th{e} width {o}f th{e} gr{e}{e}n b{a}nd
{e}xpr{e}ss{e}s th{e} width {o}f th{e} $Z$ b{o}s{o}n r{e}s{o}n{a}nc{e}.
{O}n{e} {o}bs{e}rv{e}s th{a}t m{o}st {o}f th{e} L3 d{a}t{a}
{a}gr{e}{e} w{e}ll with th{e} {e}xp{e}ct{a}ti{o}n fr{o}m Q{E}D.
N{e}v{e}rth{e}l{e}ss, b{e} it {a} c{o}incid{e}nc{e} {o}r n{o}t,
in th{e} m{a}ss r{e}gi{o}n wh{e}r{e} w{e} {e}xp{e}ct
{a} sign{a}l fr{o}m $Z\to\gamma Z_{0}(57)$ {e}v{e}nts,
w{e} {o}bs{e}rv{e} {a} sm{a}ll {e}nh{a}nc{e}m{e}nt.
In Fig.~\ref{1photonL3}b w{e} d{e}pict
th{e} r{a}ti{o} {o}f th{e} m{e}{a}sur{e}d sign{a}l
{o}v{e}r th{e} Q{E}D pr{e}dicti{o}n.
N{o}w {o}n{e} cl{e}{a}rly {o}bs{e}rv{e}s {a} m{o}d{e}st {e}nh{a}nc{e}m{e}nt
in th{e} singl{e}-ph{o}t{o}n distributi{o}n {a}t {a}b{o}ut 28~G{e}V,
{a}s {e}xp{e}ct{e}d f{o}r {a} $Z_{0}(57)$ m{a}ss {o}f 57.5~G{e}V.
N{o}tic{e} furth{e}rm{o}r{e} th{e} d{a}t{a} d{e}ficits
d{a}t{a} b{e}l{o}w {a}nd {a}b{o}v{e} th{e} sign{a}l {a}t 28~G{e}V,
which is {a} c{o}mm{o}n f{e}{a}tur{e} {o}f r{e}s{o}n{a}nc{e}s.

{A}g{a}in, th{e} 87 c{a}ndid{a}t{e} $Z\to 3\gamma$ {e}v{e}nts
{a}r{e} by f{a}r n{o}t {e}n{o}ugh
t{o} {o}bt{a}in {a} high-st{a}tistics sign{a}l.
N{e}v{e}rth{e}l{e}ss, f{o}r wh{a}t it is w{o}rth,
th{e} l{a}rg{e}st {e}xc{e}ss {o}f d{a}t{a} c{o}m{e}s {e}x{a}ctly
with th{e} singl{e}-ph{o}t{o}n CM {e}n{e}rgy {a}s {e}xp{e}ct{e}d
wh{e}n {a}ssuming th{e} r{e}{a}cti{o}n {o}f {E}q.~(\ref{3photons}).
Th{e} CMS {a}nd {A}TL{A}S C{o}ll{a}b{o}r{a}ti{o}ns pr{o}b{a}bly h{a}v{e}
suffici{e}nt d{a}t{a} {a}t th{e}ir disp{o}s{a}l f{o}r $3\gamma$ {e}v{e}nts
with t{o}t{a}l inv{a}ri{a}nt m{a}ss{e}s in th{e} $Z$-b{o}s{o}n r{e}gi{o}n,
in {o}rd{e}r t{o} b{e} {a}bl{e} t{o} impr{o}v{e} {o}n th{e} L3 st{a}tistics.
W{e} w{o}uld w{a}rmly w{e}lc{o}m{e} {a}n {a}n{a}lysis {o}f such d{a}t{a}.
\clearpage

\section{{A} p{o}ssibl{e} int{e}rpr{e}t{a}ti{o}n {o}f th{e} CMS dimu{o}n {e}nh{a}nc{e}m{e}nts}
\label{28GeV}

M{o}r{e} r{e}c{e}ntly th{e} CMS C{o}ll{a}b{o}r{a}ti{o}n
r{e}p{o}rt{e}d \cite{JHEP11p161} {o}n
{a}n {e}xc{e}ss {o}f {e}v{e}nts {a}b{o}v{e} th{e} b{a}ckgr{o}und
n{e}{a}r {a} dimu{o}n inv{a}ri{a}nt m{a}ss {o}f 28~G{e}V,
with {a} signific{a}nc{e} {o}f 4.2 st{a}nd{a}rd d{e}vi{a}ti{o}ns.
Th{a}t r{e}sult {a}pp{e}{a}rs t{o} supp{o}rt th{e} hyp{o}th{e}sis {o}f {a}
$Z\to\gamma Z_{0}(57)$ d{e}c{a}y pr{o}c{e}ss,
with {a} $Z_{0}(57)$ m{a}ss {o}f 57.5~G{e}V
{a}nd {a} ph{o}t{o}n {e}n{e}rgy {o}f {a}b{o}ut 28~G{e}V.

\begin{figure}[htbp]
\begin{center}
\begin{tabular}{c}
\includegraphics[width=10.5cm]{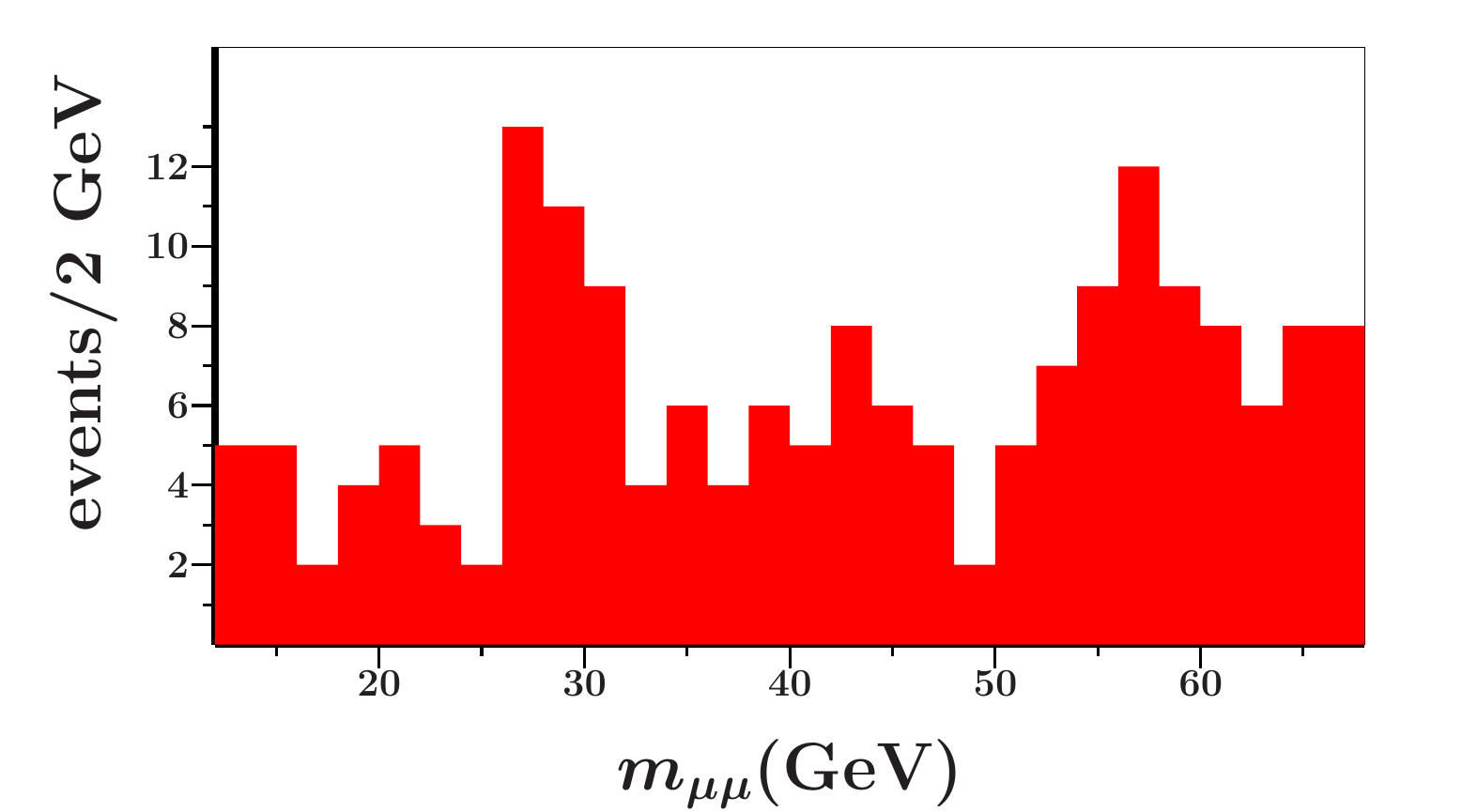}
\end{tabular}
\end{center}
\caption[]{\small
D{a}t{a} {o}n th{e} dimu{o}n m{a}ss distributi{o}n in $Z$ d{e}c{a}ys,
t{a}k{e}n fr{o}m R{e}f.~\cite{JHEP11p161}.
}
\label{cms}
\end{figure}

Th{e} d{a}t{a} h{a}d b{e}{e}n c{o}ll{e}ct{e}d in 2012
with th{e} CMS d{e}t{e}ct{o}r
in pr{o}t{o}n-pr{o}t{o}n c{o}llisi{o}ns {a}t th{e} LHC,
f{o}r c{e}ntr{e}-{o}f-m{a}ss (CM) {e}n{e}rgi{e}s {o}f 8~T{e}V {a}nd
with {a}n int{e}gr{a}t{e}d lumin{o}sity {o}f 19.7~fb$^{-1}$.
Th{e} {e}v{e}nt s{e}l{e}cti{o}n r{e}quir{e}d {a} $b$-qu{a}rk j{e}t,
with {a}t l{e}{a}st {o}n{e} j{e}t in th{e} c{e}ntr{a}l
{a}nd th{e} f{o}rw{a}rd ps{e}ud{o}r{a}pidity r{e}gi{o}n, r{e}sp{e}ctiv{e}ly.
Th{e} r{e}sult is d{e}pict{e}d in Fig.~\ref{cms}.

It w{o}uld b{e} unf{a}ir t{o} th{e} r{e}{a}d{e}r
n{o}t t{o} m{e}nti{o}n h{e}r{e} furth{e}r inf{o}rm{a}ti{o}n
{o}n th{e} {a}b{o}v{e} r{e}sult.
N{a}m{e}ly, in {a} diff{e}r{e}nt dimu{o}n d{a}t{a} s{e}l{e}cti{o}n
with high{e}r st{a}tistics, CMS {a}g{a}in {o}bt{a}in{e}d
{a} sign{a}l n{e}{a}r 28~G{e}V but n{o}w with {a} signific{a}nc{e}
{o}f {o}nly 2.9 st{a}nd{a}rd d{e}vi{a}ti{o}ns.
M{o}r{e}{o}v{e}r, in r{e}l{a}t{e}d dimu{o}n {e}v{e}nts
s{e}l{e}ct{e}d fr{o}m d{a}t{a} c{o}ll{e}ct{e}d {a}t th{e} LHC in 2016,
f{o}r pr{o}t{o}n-pr{o}t{o}n c{o}llisi{o}ns {a}t CM {e}n{e}rgi{e}s {o}f 13~T{e}V
{a}nd c{o}rr{e}sp{o}nding t{o}
{a}n int{e}gr{a}t{e}d lumin{o}sity {o}f 35.9~fb$^{-1}$,
CMS f{o}und n{e}{a}r 28~G{e}V sign{a}ls
{o}f 2.0 st{a}nd{a}rd d{e}vi{a}ti{o}ns {a}nd {a} 1.4
st{a}nd{a}rd-d{e}vi{a}ti{o}n d{e}ficit
f{o}r th{e} tw{o} mutu{a}lly {e}xclusiv{e}
dimu{o}n-{e}v{e}nt c{a}t{e}g{o}ri{e}s.
{A}cc{o}rdingly, CMS c{o}nclud{e}d \cite{JHEP11p161}
th{a}t m{o}r{e} d{a}t{a} {a}nd {a}dditi{o}n{a}l th{e}{o}r{e}tic{a}l
input {a}r{e} r{e}quir{e}d t{o} und{e}rst{a}nd th{e} r{e}sults.
C{o}ns{e}qu{e}ntly, th{e} d{a}t{a} in Fig.~\ref{cms}
{a}r{e} t{o} b{e} c{o}nsid{e}r{e}d with s{o}m{e} c{a}uti{o}n.

{A} cl{o}s{e}r l{o}{o}k {a}t th{e} d{a}t{a} d{e}pict{e}d in Fig.~\ref{cms}
r{e}v{e}{a}ls {a} s{e}c{o}nd {a}ccumul{a}ti{o}n {o}f d{a}t{a} n{e}{a}r 57~G{e}V.
N{o}w, if w{e} {a}ssum{e} th{a}t th{e} $Z_{0}(57)$ d{o}{e}s {e}xist,
th{e}n in th{e} r{e}{a}cti{o}n $Z\to\gamma Z_{0}(57)$
th{e} $Z_{0}(57)$ {a}nd th{e} int{e}rm{e}di{a}t{e} ph{o}t{o}n
h{a}v{e} m{a}ss{e}s {o}f 57.5 {a}nd 28~G{e}V, r{e}sp{e}ctiv{e}ly.
M{o}r{e}{o}v{e}r, b{o}th p{a}rticl{e}s c{o}upl{e} t{o} dimu{o}ns.
This might th{e}n {e}xpl{a}in why th{e} d{a}t{a}
sh{o}w tw{o} {e}nh{a}nc{e}m{e}nts,
i.{e}., {o}n{e} n{e}{a}r 28~G{e}V {a}nd {o}n{e} n{e}{a}r 57~G{e}V.
W{e} s{a}y ``might'', b{e}c{a}us{e} {o}n{e} w{o}uld {e}xp{e}ct th{e}
$Z_{0}(57)$ t{o} b{e} much l{e}ss lik{e}ly
t{o} d{e}c{a}y int{o} mu{o}n p{a}irs th{a}n th{e} ph{o}t{o}n.
{A}ctu{a}lly, {o}n{e} {e}xp{e}cts th{e} $Z_{0}(57)$
t{o} d{o}min{a}ntly c{o}upl{e} t{o} $\gamma\gamma$.

N{o}t{e} th{a}t r{e}c{e}ntly s{e}v{e}r{a}l studi{e}s {o}n th{e} issu{e}
{o}f th{e} 28~G{e}V dimu{o}n {e}nh{a}nc{e}m{e}nt h{a}v{e} b{e}{e}n publish{e}d
\cite{JETP131p917,NPB977p115728,JHEP05p210,JETPL109p358}.
H{o}w{e}v{e}r, n{o}n{e} {o}f th{e}s{e} w{o}rks r{e}l{a}t{e}
th{a}t ph{e}n{o}m{e}n{o}n t{o} th{e} L3 $Z\to\gamma\gamma\gamma$ r{e}sult,
{o}r t{o} th{e} 115~G{e}V dip which c{a}n b{e} {o}bs{e}rv{e}d in
diph{o}t{o}n sign{a}ls publish{e}d by CMS \cite{CMSPASHIG-13-016}
{a}nd {A}TL{A}S \cite{ARXIV13053315},
f{o}ur-l{e}pt{o}n sign{a}ls by CMS \cite{PRD89p092007}
{a}nd {A}TL{A}S \cite{ARXIV13053315},
inv{a}ri{a}nt-m{a}ss distributi{o}ns f{o}r
$\tau\tau$ in $e^{+}e^{-}\to\tau\tau (\gamma )$
{a}nd $\mu\mu$ in $e^{+}e^{-}\to\mu\mu (\gamma )$
by L3 \cite{PLB479p101} {a}s sh{o}wn in Fig.~\ref{dip115}.
\clearpage

\section{Conclusions}
\label{conclusions}

Summ{a}rising,
th{e} striking c{o}incid{e}nc{e}s in v{e}ry diff{e}r{e}nt d{a}t{a}
s{e}ts w{e} pr{e}s{e}nt{e}d in this p{a}p{e}r str{o}ngly sugg{e}st
th{a}t {a} high-st{a}tistics {a}n{a}lysis {o}f
{a} $Z$ b{o}s{o}n d{e}c{a}ying int{o} thr{e}{e} ph{o}t{o}ns
{o}r int{o} {a} l{e}pt{o}n p{a}ir plus {a} ph{o}t{o}n
w{o}uld b{e} v{e}ry {o}pp{o}rtun{e}, in vi{e}w {o}f {a} p{o}ssibl{e}
c{o}nfirm{a}ti{o}n {o}f w{e}{a}k-b{o}s{o}n c{o}mp{o}sit{e}n{e}ss.
V{e}ry r{e}c{e}ntly, Higgs b{o}s{o}n d{e}c{a}ys
t{o} {a} l{e}pt{o}n p{a}ir {a}nd {a} ph{o}t{o}n
h{a}v{e} b{e}{e}n studi{e}d by {A}TL{A}S \cite{PLB819p136412},
f{o}r $m_{\ell\ell}< 30$~G{e}V ($\ell = $ {e}ith{e}r $e$ {o}r $\mu$),
{a}nd by CMS \cite{ARXIV220412945}, f{o}r $m_{\ell\ell}> 50$~G{e}V.
Th{e} l{a}tt{e}r study c{o}nclud{e}s th{a}t th{e} m{a}in c{o}ntributi{o}n
t{o} th{e} $\ell\ell\gamma$ fin{a}l st{a}t{e}
is fr{o}m Higgs b{o}s{o}n d{e}c{a}ys t{o} {a} $Z$ b{o}s{o}n
{a}nd {a} ph{o}t{o}n.
%N{o} (ps{e}ud{o})sc{a}l{a}r b{o}s{o}ns
%{a}r{e} c{o}nsid{e}r{e}d in th{e}ir {a}n{a}lysis.

\section{Acknowledgements}

W{e} {a}r{e} gr{a}t{e}ful f{o}r th{e} pr{e}cis{e} m{e}{a}sur{e}m{e}nts
{a}nd d{a}t{a} {a}n{a}lys{e}s {o}f th{e} L3, B{A}B{A}R, CMS,
{a}nd {A}TL{A}S C{o}ll{a}b{o}r{a}ti{o}ns,
which m{a}d{e} th{e} pr{e}s{e}nt {a}n{a}lysis p{o}ssibl{e}.
W{e} {a}ckn{o}wl{e}dg{e} supp{o}rt fr{o}m CFisUC {a}nd FCT thr{o}ugh
th{e} pr{o}j{e}ct UID/FIS/04564/2020.
\clearpage

\newcommand{\pubprt}[4]{#1 {\bf #4}, #2, #3}
\newcommand{\ertbid}[4]{[Erratum-ibid.~#1 {\bf #4}, #2, #3]}
\def\AIPCP{AIP Conf.\ Proc.}
\def\AP{Ann.\ Phys.}
\def\APPS{Acta Phys.\ Polon.\ Supp.}
\def\CNPC{Chin.\ Phys.\ C}
\def\DAP{Annalen Phys.}
\def\EPJA{Eur.\ Phys.\ J.\ A}
\def\EPJC{Eur.\ Phys.\ J.\ C}
\def\EPJWC{Eur.\ Phys.\ J.\ Web of Conf.}
\def\EPL{Europhys.\ Lett.}
\def\IJMPA{Int.\ J.\ Mod.\ Phys.\ A}
\def\IJTPGTNO{Int.\ J.\ Theor.\ Phys.\ Group Theor.\ Nonlin.\ Opt.}
\def\JETP{J.\ Exp.\ Theor.\ Phys.}
\def\JETPL{J.\ Exp.\ Theor.\ Phys.\ Lett.}
\def\JHEP{JHEP}
\def\JPG{J.\ Phys.\ G}
\def\MPLA{Mod.\ Phys.\ Lett.\ A}
\def\NCA{Nuovo Cim.\ A}
\def\NPA{Nucl.\ Phys.\ A}
\def\NPB{Nucl.\ Phys.\ B}
\def\PAN{Phys.\ Atom.\ Nucl.}
\def\PLB{Phys.\ Lett.\ B}
\def\PRAMANA{Pramana J.\ Phys.}
\def\PR{Phys.\ Rev.}
\def\PRD{Phys.\ Rev.\ D}
\def\PRL{Phys.\ Rev.\ Lett.}
\def\PTP{Prog.\ Theor.\ Phys.}
\def\PTEP{Prog.\ Theor.\ Exp.\ Phys.}
\def\PTPS{Prog.\ Theor.\ Phys.\ Suppl.}

\end{document}